\documentclass{article}

\usepackage[preprint]{neurips_2026}

\usepackage[utf8]{inputenc} 
\usepackage[T1]{fontenc}    
\usepackage{booktabs}       
\usepackage{amsfonts}       
\usepackage{nicefrac}       
\usepackage{microtype}      
\usepackage{xcolor}         
\usepackage{amsmath} 
\usepackage{graphicx} 
\usepackage{multirow} 
\usepackage{pifont}
\usepackage{threeparttable}
\usepackage{tabularx}
\usepackage{makecell}
\usepackage{fvextra}
\usepackage{graphicx}
\usepackage{booktabs}
\usepackage{footnote}
\usepackage{xurl}      
\usepackage{hyperref}  
\usepackage{threeparttable}

\title{Decoupling Conversational Dynamics in Full-Duplex Spoken Models through Reinforcement Learning}
\author{%
  Yuxin Li\textsuperscript{1}  \quad Donghang Wu\textsuperscript{1} \quad Guan-Ting Lin\textsuperscript{2} \quad  Hung-yi Lee\textsuperscript{2} \quad Chengwei Qin\textsuperscript{3} \\[0.2em] \textbf{Zhehuai Chen}\textsuperscript{4} \quad \textbf{Chen Chen}\textsuperscript{4}\\[0.4em]
  \textsuperscript{1}Nanyang Technological University\quad \textsuperscript{2}National Taiwan University \\ \textsuperscript{3}The Hong Kong University of Science and Technology \quad \textsuperscript{4}NVIDIA 
}

\begin{document}

\maketitle

\begin{abstract}
Recent full-duplex spoken dialogue models have demonstrated compelling progress toward human-like interaction, enabling agents to respond with low latency, produce backchannels, and handle user barge-ins. Yet these improvements in conversational dynamics often come with weaker reasoning and instruction-following abilities, revealing a potential tension between interactive dynamics and intelligence capability. In this paper, we argue that such an \textbf{intelligence--dynamics} trade-off is not fundamental: conversational dynamics can instead be learned as a separate real-time decision policy from human dialogue data. To this end, we propose DuplexPO, a reinforcement learning (RL) framework that decouples when to speak from what to say. It preserves the semantic response capability of an instruction-tuned assistant, while optimizing its temporal interaction behavior over selected high-impact windows from long human conversations. To quantitatively optimize these dynamics, we formulate the Factorized Conversational Dynamics Reward (FCDR) to enable fine-grained temporal credit assignment for turn initiation, backchanneling, yielding, and regularized participation. The policy is then optimized with a GRPO-style objective. Experiments show that DuplexPO substantially improves full-duplex behaviors, including timely backchannels, smooth turn-taking, and barge-in handling, while maintaining strong reasoning and instruction-following performance. Moreover, improvements in dynamics-oriented metrics are reflected in better user experience, suggesting that optimizing conversational timing as a standalone objective can promote more natural full-duplex interaction.
\end{abstract}

\section{Introduction} 

Speech is widely regarded as the most natural modalities for human-computer interaction, making spoken dialogue systems a long-standing focus in both academia and industry. In recent years, advances in neural modeling and large-scale data have substantially accelerated progress in this area \citep{slm_survey1}. In particular, end-to-end spoken dialogue models have emerged as an increasingly important paradigm for building interactive and helpful voice agents \citep{dgslm}. \par

More recently, full-duplex spoken dialogue models have gained increasing attention in the research community due to their ability to support simultaneous listening and speaking \citep{moshi, salmonnomni}. By allowing the model to process incoming user speech while generating responses, this paradigm enables real-time agent behaviors such as handling user barge-ins and producing timely backchannels. As a result, full-duplex modeling offers a more responsive and human-like conversational experience compared to traditional turn-based approaches \citep{chronological}. \par

However, such gains in conversational dynamics may come at the expense of model intelligence: full-duplex models perform substantially worse on instruction-following and reasoning benchmarks than their half-duplex counterparts \citep{salmduplex, voiceagentbench}, suggesting a potential \textbf{intelligence--dynamics} trade-off. We analyze this trade-off from two complementary perspectives. First, at the modeling level, deep logical deduction and fine-grained temporal coordination impose different demands on autoregressive generation. Strong reasoning often requires long and structured decoding trajectories that support multi-step inference \citep{cot, audioreasoner}, while natural full-duplex interaction requires the model to continuously process fragmented, rapidly changing contexts and make local real-time decisions about whether to speak, pause, or remain silent. Such temporal control may bias the model toward short-range interaction states, making it harder to preserve the stable long-range dependencies needed for complex reasoning. Second, at the data level, training conversational behaviors often relies on human dialogue corpora such as Fisher~\citep{cieri2004fisher} during supervised fine-tuning (SFT). Although such data are useful for learning interaction patterns, they are dominated by casual or non-goal-oriented exchanges and are therefore not necessarily aligned with the goal of building a helpful, instruction-following assistant. \par

Based on these analyses, we argue that the above conflict largely stems from coupling ``\textit{what to say}'' and ``\textit{when to speak}'' in a single learning objective. These decisions are fundamentally different: the former concerns semantic content generation and is the primary target of supervised instruction tuning, while the latter governs temporal behaviors essential to full-duplex interaction, such as turn-taking, backchanneling, and barge-in handling \citep{fullduplex-bench}. 
This distinction suggests that full-duplex conversational dynamics need not be relearned by imitating complete natural dialogue responses, which are dominated by casual, non-goal-oriented exchanges that diverge from assistant-style interaction. Instead, an instruction-tuned spoken language model can preserve its semantic competence while selectively adapting its policy at the moments where timing decisions about turn-taking, backchanneling, and yielding actually matter. 
This motivates our central hypothesis: by optimizing only real-time floor-control decisions, a full-duplex spoken language model can improve conversational dynamics while preserving its instruction-following and reasoning abilities.
This follows the broader principle of targeted speech-specific alignment, where auxiliary speech capabilities are learned without replacing the model's general language capability \citep{chen2025audio}. \par

To operationalize this separation, we propose DuplexPO, an RL framework that treats conversational dynamics as an independent optimization target for full-duplex spoken dialogue models. Concretely, DuplexPO samples dynamic-critical windows from long multi-turn human conversations, focusing optimization on informative full-duplex events such as turn transitions, backchannels, and barge-ins. Since trajectories are sampled from the model itself, the policy is improved under its own response distribution rather than forced to imitate heterogeneous conversational corpora. 
It then computes a Factorized Conversational Dynamics Reward (FCDR) from the sampled windows and uses it as the reward signal for RL. Importantly, we show that optimizing these reward maps translates into an improved user experience: higher dynamics scores lead to interactions that users perceive as more natural, responsive, and temporally well-coordinated. 
Finally, the GRPO-style objective regularizes the updated policy toward the reference SFT model, limiting policy drift and retaining the semantic behaviors acquired during instruction tuning \citep{schulman2017proximal, ouyang2022training, shao2024deepseekmath}.
\par

Empirically, DuplexPO improves turn-taking, backchanneling, and barge-in handling across multiple conversational-dynamics evaluations, while maintaining comparable performance on factual QA, instruction following, speech understanding, and reasoning benchmarks. We also provide a demo page at \url{https://liyuxin44.github.io/DuplexPO/}.\par

\section{Related Work}
\subsection{Full-Duplex Dialogue: Systems vs. Models}
Traditional spoken dialogue systems usually decompose real-time interaction into recognition, dialogue management, incremental processing, and synthesis modules \citep{streaming_asr, he2019streaming, schlangen2011general, Incremental_Processing, skerry2018towards}. Full-duplexity is therefore best understood as a functional property rather than a single architecture. A system can obtain it through external orchestration, such as VAD or dialogue-management control \citep{vad_analysis, semanticvad, zhang2025llm}, while a model can internalize it as part of its own generation process \citep{soulxduplug, veluri2024beyond, wang2024full}. In system-level designs, auxiliary control can determine \textit{when to speak} while the core model focuses on response generation.

Model-level full duplex follows a different formulation. Models such as Moshi, SALMONN-omni, SALM-Duplex, and related latent-reasoning variants expose turn-taking, backchanneling, and barge-in behavior directly to the model \citep{moshi, salmonnomni, salmduplex, flair}. Recent full-duplex benchmarks similarly treat these behaviors as explicit model capabilities rather than peripheral interface features \citep{fullduplex-bench, lin2026full_15, lin2026full}. This paper therefore focuses on model-level full-duplex dialogue, where the key challenge is not only generating an appropriate answer, but learning a temporally coordinated policy for deciding when that answer should enter, pause, or leave the shared speech stream.

\subsection{Reinforcement Learning for Dialogue Agents}

RL is well suited to dialogue policy optimization, where actions often have delayed effects beyond next-token likelihood \citep{sutton2018reinforcement}. Prior work has applied RL to task-oriented dialogue, turn-taking, conversational reasoning, retrieval-augmented QA, and AI-feedback-based response optimization \citep{raux2012optimizing, khouzaimi2016reinforcement, speakrl, chatr1, arora2026optimizing, dialogxpert}. This direction has also begun to reach end-to-end spoken language models. Align-SLM uses AI-feedback preference optimization to improve semantic coherence in textless SLMs, while user-interaction alignment constructs large-scale preference pairs from raw multi-turn speech conversations for full-duplex speech-to-speech models \citep{lin2025align, wu2025aligning}. These studies show that reward-based learning is useful when the desired behavior is sequential, context-dependent, or only weakly specified by supervised targets.

Full-duplex dynamics pushes this view into a more local temporal regime. Yielding after user barge-in, producing a brief backchannel, or delaying a turn start are decisions whose quality depends on a narrow acoustic and conversational context, yet whole-dialogue objectives can blur their credit assignment \citep{dream}. Recent spoken-dialogue RL and reward-modeling work has begun to explicitly optimize interaction behaviors such as turn-taking and backchanneling by learning dialogue-level policies or reward models over entire conversations \citep{chen2025reinforcement, chen2026dual}. In contrast, our approach focuses on fine-grained temporal credit assignment, restricting policy updates to short, interaction-critical windows rather than optimizing over full dialogues. ASPIRin shows that raw-token RL for full-duplex timing can degrade semantic quality, and addresses this by projecting the action space into active-speech versus inactive-silence states before GRPO-style optimization \citep{hsiao2026aspirin}. DuplexPO instead keeps the original action space and changes the optimization unit, updating only dynamic-critical windows rather than the full dialogue.

\begin{figure}[t]
    \centering
    \includegraphics[width=1.0\linewidth]{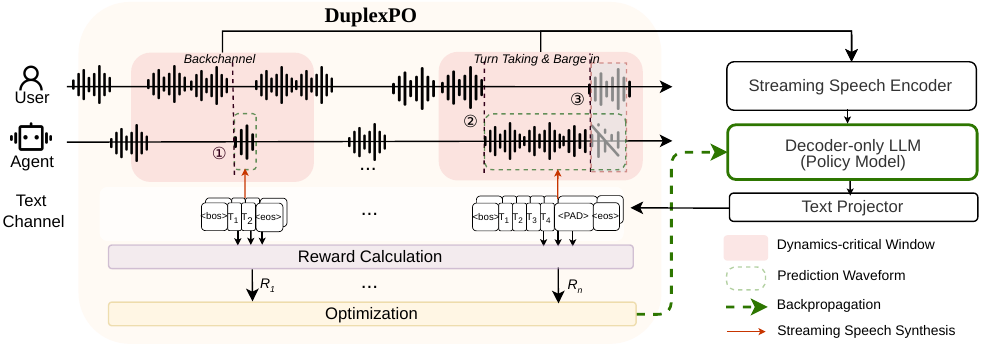}
    \caption{Overview of the DuplexPO for full-duplex spoken dialogue models. The red shadows indicate dynamics-critical windows that may include important conversational behaviors, such as: \textcircled{1} backchanneling, \textcircled{2} turn-taking, and \textcircled{3} user barge-in. For each window, the policy model generates multiple rollouts for reward calculation and is optimized with RL.}
    \label{fig:model}
\vspace{-15pt}
\end{figure}

\section{Methodology}
\subsection{Overview}
We propose DuplexPO, a policy optimization method that decouples when and how to engage in conversation (turn-taking, backchanneling, yielding to user barge-ins) from what to say (semantic content). DuplexPO improves full-duplex behavior while preserving the model's instruction-following and reasoning capabilities. As shown in Figure~\ref{fig:model}, the framework consists of three main components. First, Dynamic-critical Window Sampling selects local windows around annotated agent speaking events and restricts policy optimization to these regions. Second, the Factorized Conversational Dynamics Reward (FCDR). Third, group-based policy optimization. Appendix~\ref{app:model-analysis} provides a token-level example showing how DuplexPO reshapes boundary-control decisions.

\subsection{Problem Formulation}
We formulate full-duplex conversational dynamics learning as frame-level policy optimization over streamed dialogue. Let $x_{1:T}$ denote the user audio stream and $y_{1:T}$ denote the agent's real-time interaction decisions on a uniform temporal grid with frame duration $\Delta$. At each frame, the policy predicts
\begin{equation}
\pi_\theta(y_t \mid y_{<t}, x_{\leq t}),
\label{eq:duplex-policy}
\end{equation}
where $x_{\leq t}$ is the causal user context and $y_t$ represents the agent's interaction state, such as silence, speaking, turn initiation, or yielding.

Instead of optimizing over entire conversations, DuplexPO operates on a set of local dynamics-critical windows $\mathcal{W}=\{W_i\}_{i=1}^{M}$ extracted from long human conversations. Each window contains a short temporal region around an interaction-critical event, together with its reference speaking interval and event metadata, as defined in subsection \ref{sec:window_sampling}. For a window $W_i$ with temporal support $[s_i,e_i)$, the policy samples actions only inside the window while conditioning on the teacher-forced history before $s_i$. The sampled sub-trajectory is evaluated by a window-level reward:
\begin{equation}
R_i = R(W_i, y_{[s_i,e_i)}),
\label{eq:window-reward}
\end{equation}
where $y_{[s_i,e_i)}$ denotes the agent's sampled decisions within the window.

\subsection{Dynamics-Critical Window Sampling}
\label{sec:window_sampling}

Prior study suggests that conversational turn transfer is often determined by local timing and prosodic cues near possible response points, rather than by evidence uniformly distributed across the entire dialogue \citep{ward2000prosodic, sacks1974simplest}. Therefore, DuplexPO does not optimize the policy over full conversations. Instead, it selects short but important local windows, such as turn transitions, backchannels, and user barge-ins, where real-time speaking decisions are most critical.

Let $(a_i,b_i)$ denote the start and end times of the $i$-th annotated human agent segment, and let $c_i$ denote its event metadata, including whether the segment is a backchannel and whether the corresponding window contains user barge-in. Each window is defined as
\begin{equation}
W_i = (s_i, e_i, g_i, h_i, c_i),
\end{equation}
where $[g_i,h_i)$ is the discretized reference speaking interval and $[s_i,e_i)$ is the sampled window. The reference interval is
\begin{equation}
g_i=\left\lfloor a_i/\Delta \right\rfloor, \qquad
h_i=\left\lfloor b_i/\Delta \right\rfloor .
\end{equation}

The sampled window includes a lead time $L$ before the annotated segment and a buffer time $B$ after it:
\begin{equation}
s_i=\max\!\left(e_{i-1},\,\left\lfloor (a_i-L)/\Delta \right\rfloor\right),
\qquad
\bar{e}_i=\left\lfloor (b_i+B)/\Delta \right\rfloor .
\end{equation}
For the first window, the previous boundary term is omitted. We further apply boundary clipping to avoid overlap between neighboring events:
\begin{equation}
e_i=\min\!\left(\bar{e}_i,\,
\left\lfloor (a_{i+1}-L)/\Delta \right\rfloor,\,
T\right),
\end{equation}
where the next-segment boundary is omitted for the final segment.

Within each window, the history before $s_i$ is teacher-forced, and the policy is optimized only on sampled actions in $[s_i,e_i)$.

\subsection{Factorized Conversational Dynamics Reward (FCDR)}
\label{sec:FCDR}

Although ORISE~\citep{chen2025reinforcement} shows that rollout-level rewards can improve spoken interaction, its sequence-level assignment provides coarse supervision, making it difficult to attribute success or failure to specific real-time decisions. We therefore introduce a temporally shaped reward that assigns fine-grained supervision at the event level. The proposed reward design is motivated by findings from human turn-taking and backchannel behavior in conversational interaction \citep{sacks1974simplest, stivers2009universals, levinson2015timing}.

For a sampled continuation in window $W_i$, let $\hat{\mathcal{S}}_i\subseteq[s_i,e_i]$ denote the frames where the agent speaks. If $\hat{\mathcal{S}}_i\neq\emptyset$, the predicted onset is $\hat{g}_i=\min \hat{\mathcal{S}}_i$, and the onset delay is
\begin{equation}
\tau_i = \Delta(\hat{g}_i-g_i),
\end{equation}
where $\tau_i<0$ means the agent starts too early and $\tau_i>0$ means it starts late. We derive three binary masks from the event metadata $c_i$: $m_{\mathrm{on}}^i$, $m_{\mathrm{bc}}^i$, and $m_{\mathrm{off}}^i$, indicating whether onset timing, backchannel timing, and barge-in yielding are active for the current window.

The final FCDR is a weighted sum of factorized event-level components:
\begin{equation}
R_{\mathrm{FCD}}(W_i,\hat{\mathcal{S}}_i)
=
\lambda_{\mathrm{on}} m_{\mathrm{on}}^i R_{\mathrm{on}}^i
+
\lambda_{\mathrm{bc}} m_{\mathrm{bc}}^i R_{\mathrm{bc}}^i
+
\lambda_{\mathrm{off}} m_{\mathrm{off}}^i R_{\mathrm{off}}^i
+
\lambda_{\mathrm{reg}} R_{\mathrm{reg}}^i .
\end{equation}

The component definitions are summarized in Table~\ref{tab:DuplexPO-rewards}. This factorized design assigns credit to local timing decisions while keeping the reward interpretable across different conversational-dynamics events. Figure~\ref{fig:reward_component_relationships} in Appendix provides an empirical analysis of the reward components. We also compare FCDR with neural reward model alternatives in Appendix~\ref{app:NRM}, including designs based on temporal state prediction and window-level dynamics scoring.

\begin{table}[h]
\centering
\caption{
FCDR components. Each component is computed within a dynamics-critical window $W_i$. Event-specific masks determine which components are active. Here $\mathcal{B}_i=[g_i,h_i]$ denotes the annotated backchannel interval, $d(t,\mathcal{B}_i)$ denotes the frame distance from $t$ to $\mathcal{B}_i$, $\ell_i$ denotes the duration of agent speech after user barge-in, and $e_{i,k}$ denotes predefined undesirable interactions.
}
\footnotesize
\label{tab:DuplexPO-rewards}
\begin{tabular}{lccc}
\toprule
Component & Target behavior & Functional form & Range \\
\midrule

$R_{\mathrm{on}}^i$
&
Turn initiation
&
$\mathbb{I}[\hat{\mathcal{S}}_i\neq\emptyset]
\exp\!\left(
-\dfrac{\tau_i^2}{2\sigma_{\mathrm{on}}(\tau_i)^2}
\right)$
&
$[0,1]$
\\[0.8em]

$R_{\mathrm{bc}}^i$
&
Backchanneling
&
$\mathbb{I}[\hat{\mathcal{S}}_i\neq\emptyset]
\max\limits_{t\in\hat{\mathcal{S}}_i}
\left(
\mathbb{I}[t\in\mathcal{B}_i]
+
\mathbb{I}[t\notin\mathcal{B}_i]e^{-\alpha d(t,\mathcal{B}_i)}
\right)$
&
$[0,1]$
\\[1.1em]

$R_{\mathrm{off}}^i$
&
Yielding after barge-in
&
$-\mathrm{clip}
\left(
\dfrac{\max(0,\ell_i-\ell^{\ast})}{H_{\mathrm{off}}},
0,
1
\right)$
&
$[-1,0]$
\\[1.0em]

$R_{\mathrm{reg}}^i$
&
Pattern regularization
&
$-\mathrm{clip}
\left(
\sum_{k=1}^{K}\beta_k\mathbb{I}[e_{i,k}],
0,
1
\right)$
&
$[-1,0]$
\\

\bottomrule
\end{tabular}
\end{table}

\subsection{Group-based Optimization}

DuplexPO uses a GRPO-style objective~\citep{shao2024deepseekmath} to optimize sampled continuations within each dynamics-critical window. GRPO uses group-normalized advantages over the full set of sampled continuations, providing a denser and more stable optimization signal for dynamics-critical decisions than DPO-style preference optimization.

For window $i$, we normalize rewards across its sampled continuations to obtain a clipped group advantage:
\begin{equation}
A_{i,k} =
\mathrm{clip}\!\left(
\frac{r_{i,k}-\mu_i}{\sigma_i+\epsilon}, -5, 5
\right),
\end{equation}
where $r_{i,k}$ is the reward of the $k$-th continuation, and $\mu_i,\sigma_i$ are the reward mean and standard deviation within the same window group. We then apply a policy-gradient loss only to sampled tokens inside the window, with a KL penalty toward the rollout behavior policy:
\begin{equation}
\mathcal{L}_{\mathrm{GRPO}}
=
\mathcal{L}_{\mathrm{policy}}
+
\beta \mathcal{L}_{\mathrm{KL}} .
\end{equation}

We also implement a DPO-style baseline by constructing pairwise preferences among continuations from the same window, and report the ablation in Appendix \ref{app:opt-ablation}.

\section{Data and Model Training}
\label{sec:data_and_training}
\subsection{Conversational Dynamics-Aware Dialogue Data Generation}
\label{sec:dynamic-data}
Fisher~\citep{cieri2004fisher} and Seamless-Naturalistic-HQ
(Seamless)~\citep{agrawal2025seamless} provide natural two-party
conversations with word-level timing, making them suitable sources for learning conversational dynamics. The two corpora cover complementary interaction regimes: Fisher is closer to symmetric peer-to-peer conversation, whereas Seamless contains more role-asymmetric, question-answer-oriented interactions. This combination enables us to test whether DuplexPO can handle both balanced conversational exchange and assistant-like spoken interaction. All Fisher and Seamless evaluations use held-out conversations that are split before reconstruction and window annotation; the training conversations, derived windows, and timing annotations are disjoint from the evaluation conversations and evaluation annotations. Reconstruction, filtering, and sampling details for data preprocessing are given in Appendix~\ref{app:data-details}.

\subsection{Training Pipeline}
The training pipeline consists of four stages: pre-training, SFT, RL with DuplexPO, and speech-synthesis SFT. The pre-training and SFT stages follow the setup of Nemotron-VoiceChat~\citep{voicechat} and we provide more training and data details in Appendix~\ref{app:data-details}. Then, DuplexPO is applied to optimize conversational dynamics with explicit rewards over dynamics-critical windows. Finally, in speech-synthesis SFT, we freeze all parameters except the speech generation module and train the streaming TTS model on the same speech dataset~\citep{cosyvoice2}. The implementation, optimization, and stage-wise hyperparameters are in Appendix~\ref{app:training-details}.

\section{Evaluation and Benchmark}
\label{sec:evaluation}

We evaluate Spoken Dialogue Language Models (SDLMs) from two complementary aspects, namely conversational dynamics and model intelligence. The former assesses full-duplex conversational dynamics, such as turn-taking and backchanneling. The latter examines general capabilities, including factual knowledge, instruction following, speech understanding, and reasoning.

\subsection{Conversational Dynamics Evaluation}
\label{sec:dynamics-eval}

Conversational dynamics are evaluated on Fisher~\citep{cieri2004fisher}, Seamless~\citep{agrawal2025seamless}, and Full-Duplex-Bench v3 (FDB-v3)~\citep{lin2026full}. For each conversation, we evaluate all non-overlapping windows spanning the full dialogue. On Fisher and Seamless, we report window-level initiation rate, onset MAE, and yield rate for both full-turn and backchannel events, measuring \texttt{<BOS>} occurrence, \texttt{<BOS>} timing error, and timely \texttt{<EOS>} emission, respectively. Here, \texttt{<BOS>} and \texttt{<EOS>} denote the beginning and end of the agent's speaking turn, respectively. Onset MAE is computed only for initiated events. FDB-v3 tests robustness to premature agent speech during disfluent user turns and user re-entry. We report Turn-taking Latency, Voiced Interrupt Rate (VIR), and Yield Rate. VIR measures the percentage of agent speech onsets that overlap with voiced user speech. We exclude onsets during internal user pauses and within the final 0.5s of a voiced user segment, where overlap may reflect turn-transition ambiguity rather than interruption. Yield Rate measures whether the agent releases the floor before the user resumes speaking.

Following prior pairwise preference evaluation protocols for open-ended dialogue models~\citep{zheng2023judging, chiang2024chatbot}, we conduct a blinded matched-window LLM-as-a-Judge evaluation with Gemini 3.0 Pro\footnote{\url{https://gemini.google.com/}} to compare DuplexPO with the \emph{SFT Baseline} at the conversation level. For each aligned dialogue, the judge is given the shared user timeline and two anonymized candidate model timelines. To fit the judge's context budget and control evaluation cost, we sample a subset of windows per dialogue, using the same window indices for both models to ensure matched conversational contexts. We concatenate each model's responses over these windows into a single conversation-level judge input. The judge selects the candidate with more natural conversational dynamics and reports dimension-level preferences for turn-taking, backchanneling, and user barge-in handling. The full prompts, and per-dimension results are provided in Appendix~\ref{app:llm-pairwise-judge}.


\subsection{Model Intelligence Evaluation}
\label{sec:intelligence-eval}

To verify that conversational dynamics optimization preserves task-level capability, we evaluate SDLMs on factual knowledge, instruction following, speech understanding, and reasoning benchmarks. Factual knowledge is measured by accuracy on four quick QA sets: Llama Questions~\citep{llamaq}, WebQuestions~\citep{webquestions}, TriviaQA~\citep{triviaqa}, and SDQA~\citep{faisal2021sd}. Instruction following is evaluated with AlpacaEval and CommonEval from VoiceBench~\citep{voicebench}; the latter uses real human speech and better reflects spoken instruction-following scenarios. Following the official VoiceBench protocol, open-ended responses are scored by a GPT judge on a 1--5 scale. Speech understanding and reasoning are measured by multiple-choice accuracy on OpenBookQA and MMSU~\citep{mmsu}. For all benchmarks, generated speech is transcribed with Whisper large v3~\citep{whisper} before computing text-based metrics under the official VoiceBench protocol.

\begin{table}[t]
\centering
\caption{Window-level turn-taking and backchannel performance on Fisher and Seamless.
Rates (except Onset MAE) are reported in \%. Best values in each column are bolded.}
\label{tab:rule-based}
{
\renewcommand{\arraystretch}{0.9}
\resizebox{\textwidth}{!}{
\begin{tabular}{lcccccc}
\toprule
\multirow{2}{*}{Method} & \multirow{2}{*}{Dataset} & \multirow{2}{*}{Onset MAE ($\downarrow$)} & \multicolumn{2}{c}{Turn-taking} & \multicolumn{2}{c}{Backchannel} \\
\cmidrule(lr){4-7}
 &  &  &Init Rate ($\uparrow$) & Yield Rate ($\uparrow$) & Init Rate ($\uparrow$) & Yield Rate ($\uparrow$) \\
\midrule
Moshi \citep{moshi}& \multirow{5}{*}{Fisher} & 1.99 & 87.7&76.5 &83.9 &66.7 \\
PersonaPlex \citep{roy2026personaplex}& &1.50&98.6  &80.3 & 91.4&69.2 \\
\emph{SFT Baseline}& &0.98 & 97.8  &92.1 &95.7 &57.1 \\
\emph{SFT Dynamics} & & 1.14& 97.1 &78.9 &91.4 &57.1 \\
DuplexPO & &\textbf{0.69}& \textbf{100.0} &\textbf{98.7} &\textbf{97.8} &\textbf{100.0} \\
\midrule
Moshi \citep{moshi}& \multirow{5}{*}{Seamless} &2.16 &80.1  &69.0 &76.2 &71.3 \\
PersonaPlex \citep{roy2026personaplex} & & 1.77 & 82.8& 79.6&82.3 &84.0 \\
\emph{SFT Baseline} &  &1.22& 91.8 &78.5 &92.2 &79.4 \\
\emph{SFT Dynamics} &  & 1.34& 79.7& 63.0&83.2 & 63.4\\
DuplexPO & &\textbf{1.03} &\textbf{98.0} &\textbf{93.6} & \textbf{99.5}& \textbf{93.3}\\
\bottomrule
\end{tabular}
}
}
\end{table}

\begin{figure}[h]
  \centering
  \includegraphics[width=0.9\linewidth]{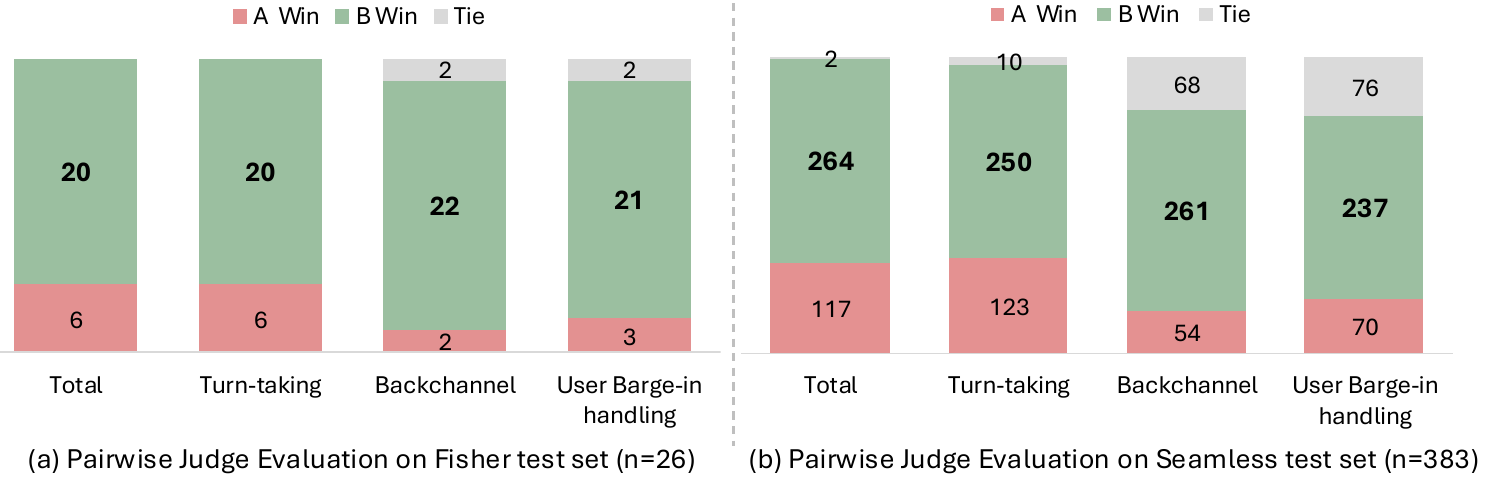}
    \caption{Conversation-level Gemini pairwise evaluation comparing the \emph{SFT Baseline}
    (A) with DuplexPO (B). The judge observes timestamps, transcripts, and aggregate dynamics
    statistics to assess turn-taking, backchanneling, and barge-in handling.}
  \label{fig:gemini_eval} 
\end{figure}

\begin{table}[t]
\centering
\caption{Turn-Taking Dynamics, VIR, and Yield Rate on FDB-v3.
Rates are reported in \%, and Latency is reported in seconds. The results of GPT-Realtime, Gemini Live 2.5, Gemini Live 3.1, Grok and Ultravox are taken from ~\citep{lin2026full}. Best values in each column are bolded.}
\label{tab:disfluency_barge_in}
{
\footnotesize
\setlength{\tabcolsep}{3.5pt}
\renewcommand{\arraystretch}{0.8}
\begin{tabular}{lcccc}
\toprule
\multirow{2}{*}{Method}
& \multicolumn{2}{c}{Turn-Taking Dynamics}
& \multirow{2}{*}{VIR ($\downarrow$)}
& \multirow{2}{*}{Yield ($\uparrow$)} \\
\cmidrule(lr){2-3}
& Turn-Take ($\uparrow$) & Latency ($\downarrow$) & & \\
\midrule
GPT-Realtime \citep{openai_gptrealtime_1_5}
& 96.0 & 2.65 &-- &  --\\
Gemini Live 2.5 \citep{google_gemini_2_5_flash_audio}
& 92.0 & 3.84 & --& -- \\
Gemini Live 3.1 \citep{google_gemini_3_1_flash_live}
& 78.0 & 1.59  &-- &  --\\
Grok \citep{x_grok_voice_agent}
& 94.0 & 3.13  & -- &  --\\
Ultravox \citep{fixie_ai_ultravox}
& 96.0 & 1.90  &-- &  --\\
Moshi~\citep{moshi}
& \textbf{100.0} & 0.44  & 12.0& 71.9 \\
PersonaPlex~\citep{roy2026personaplex}
& 98.0 & 1.41  & 15.0& 66.7 \\
\midrule
\emph{SFT Baseline}
& 99.0 & 7.33  & 8.0&64.8 \\
\emph{SFT Dynamics}
& \textbf{100.0} & 14.24 & \textbf{4.0} & 93.3 \\
DuplexPO
 &\textbf{100.0}  & \textbf{0.24}& 5.0 & \textbf{100.0} \\
\bottomrule
\end{tabular}

}
\end{table}

\section{Results and Analysis}
\subsection{Conversational Dynamics}

\begin{table}[t]
\centering
\caption{Model intelligence evaluation. QA, OpenBookQA, and
MMSU metrics are accuracy in \%; AlpacaEval and CommonEval are GPT-scores on a
1--5 scale. Baseline results are taken from \citep{salmonnomni, voicebench}.
FD denotes full duplex. LlamaQ, WebQ, TriQA, OBQA, AlpacaE, and ComE denote
Llama Questions, WebQuestions, TriviaQA, OpenBookQA, AlpacaEval, and
CommonEval.}
\label{tab:qa}
{
\renewcommand{\arraystretch}{0.9}
\resizebox{\textwidth}{!}{
\begin{tabular}{lccccccccc}
\toprule
{Method} & {FD} & {LlamaQ} & {WebQ} & {TriQA} & {SDQA} & {AlpacaE} 
& {ComE}
& {OBQA}
& {MMSU} \\
\midrule
Moshi \citep{moshi} & \ding{51} & 54.5 & 22.1 & 16.7 
& 15.6
& 2.01
& 1.60
& 25.9
& 24.0 \\
Freeze-Omni \citep{freezeomni} & \ding{51} & 56.2 & 27.9 & 28.5 
& 53.5
& 4.03 & 3.46 & 31.0 & 28.1 \\
SALMONN-omni \citep{salmonnomni} & \ding{51} & 73.6 & 43.7 & 56.0 
& -
& 3.22 & - & - & 30.0 \\
SALM-Duplex \citep{salmduplex} & \ding{51} & 51.3 & 25.0 & 16.9 & 26.0 & 2.99 & 2.50 & 39.6 & 26.3 \\
\midrule
GLM-4-Voice \citep{glm4voice} & \ding{55} & 65.7 & 37.0 & 47.5 
& 37.0
& 3.97
& 3.42 
& 53.4 & 39.8 \\
Qwen2-Audio \citep{qwen2audio} & \ding{55} & 69.7 & 45.2 & 40.3 & 35.7 & 3.74 & 3.43 & 49.5 & 35.7 \\
Kimi-Audio \citep{kimi} & \ding{55} & 68.3 & 37.3 & 51.2 
& 63.1
& 4.46
& 3.97 & 83.5 & 62.2 \\
Baichuan-Audio \citep{baichuan} & \ding{55} & 74.0 & 40.7 & 53.0 
& 45.8
& 4.41
& 4.08 & 71.7 & 53.2 \\
STITCH-R \citep{stitch} & \ding{55} & 70.0 & 50.3 & 49.6 
& - 
& 2.70 & - & - & - \\
\midrule
\emph{SFT Baseline} & \ding{51} & 72.0 &44.3 & 48.1 & 47.2 & 3.43  & 3.48 & 72.2&54.9\\
DuplexPO & \ding{51} & \textbf{75.3}& \textbf{44.5}& \textbf{49.9} & \textbf{49.8}
& \textbf{3.68}
& \textbf{3.74}
& \textbf{73.7}& \textbf{56.2}\\
\bottomrule
\end{tabular}
}
}
\end{table}

We compare DuplexPO with two SFT-only baselines. \emph{SFT Baseline} is trained without dynamics-aware dialogue data, whereas \emph{SFT Dynamics} uses the same SFT recipe as DuplexPO and includes the reconstructed dynamics-aware dialogue data.

Table~\ref{tab:rule-based} shows that DuplexPO consistently improves window-level turn-taking and backchannel behavior on both Fisher and Seamless. \emph{SFT Dynamics} shows that SFT alone is insufficient for learning robust conversational coordination from natural dynamics patterns, regressing across all dynamics metrics relative to the \emph{SFT Baseline}. DuplexPO achieves the best overall performance, with the highest initiation and yield rates for both turn-taking and backchannels, as well as the lowest onset MAE on both datasets.

Table~\ref{tab:disfluency_barge_in} examines whether these window-level gains transfer to ambiguous mid-turn regions where premature responses cause barge-ins. The results reveal a latency-interruption trade-off among existing models. Commercial models and Ultravox achieve reasonable turn-taking accuracy but incur high latency, reflecting conservative endpointing strategies that wait for extended silence. In contrast, open-source duplex models such as Moshi and PersonaPlex reduce latency through always-on listening, but suffer from substantially higher VIR due to premature interruptions. Among our ablations, \emph{SFT Dynamics} attains low VIR and high Yield Rate only by adopting long latencies, effectively trading responsiveness for caution.

DuplexPO breaks this trade-off. It preserves perfect turn-taking, achieves the highest Yield Rate and the lowest latency among all models, and keeps VIR competitively low. These results suggest that dynamics-aware RL learns coordinated floor control rather than simply biasing the agent toward eagerness or caution. By using mid-turn dynamics evidence to distinguish true completion from disfluent continuation, DuplexPO can respond promptly when the user yields while reliably yielding when the user retains the floor. The reward curve and pairwise correlations among sub-reward components are visualized in Appendix~\ref{app:reward_vis}.

Figure~\ref{fig:gemini_eval} evaluates whether these metric gains are reflected at the conversation level. The pairwise judge prefers DuplexPO in $76.9\%$ of Fisher comparisons and $69.3\%$ of non-tie Seamless comparisons. Dimension-level preferences are consistent with the automatic metrics for turn-taking, backchanneling, and user barge-in handling.

\subsection{Model Intelligence}
Table~\ref{tab:qa} addresses the second side of the proposed intelligence--dynamics trade-off. Across factual QA, instruction following, speech understanding, and reasoning, DuplexPO preserves the \emph{SFT Baseline}'s task-level capability and yields small but consistent improvements. Although the magnitude of these gains is modest, their direction is important because optimizing the real-time speaking policy does not degrade the semantic and reasoning abilities learned during SFT.

\subsection{Discussion and Visualization}

\begin{figure}[h]
  \centering
  \begin{minipage}[t]{0.48\linewidth}
    \centering
    \vspace{0pt}
    \includegraphics[width=\linewidth]{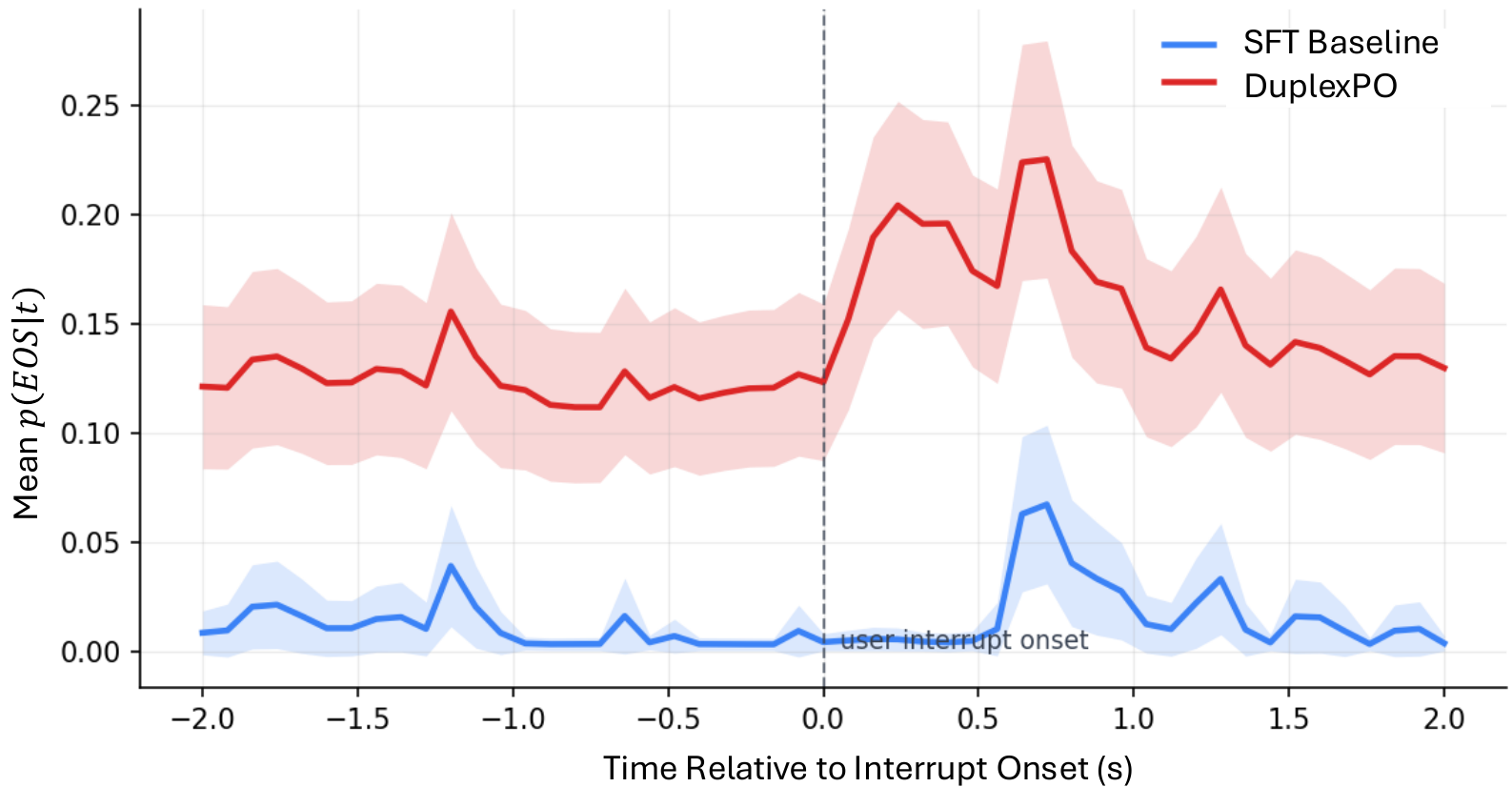}
    \small (a) \texttt{<EOS>} impulse around user interruption onset. 
  \end{minipage}
  \hfill
  \begin{minipage}[t]{0.48\linewidth}
    \centering
    \vspace{0pt}
    \small
    \begin{tabular}{lcccc}
      \toprule
      \multirow{2}{*}{Dataset} & \multicolumn{2}{c}{\texttt{<BOS>} Impulses} & \multicolumn{2}{c}{\texttt{<EOS>} Impulses} \\
      \cmidrule(lr){2-3} \cmidrule(lr){4-5}
       & SIR & SRR & SIR & SRR \\
      \midrule
      Fisher & 11.0 & 70.6 & 4.9 & 100.0 \\
      FDB-v3 & 10.8 & 71.9 & 0.9 & 100.0 \\
      \bottomrule
    \end{tabular}
    \smallskip

    \small (b) Suppressed-intent analysis results for \texttt{<BOS>} and \texttt{<EOS>} impulses. Values are reported in \%.
  \end{minipage}

  \caption{Analysis of suppressed boundary intent in \emph{SFT Baseline} and DuplexPO. (a) \texttt{<EOS>} impulse around user interruption onset on Fisher test set ($n{=}148$ events from 20 conversations). Both policies are evaluated on the SFT trajectory; shaded regions show 95\% CI. (b) Suppressed-intent analysis.}
  \label{fig:eos_and_diagnostics}
\end{figure}

We analyze how DuplexPO reshapes conversational behavior using two metrics defined in Appendix~\ref{app:sir-srr}: Suppressed Intent Rate (SIR), measuring how often latent boundary intent is suppressed, and Suppression Release Ratio (SRR), measuring whether suppressed intent is later released. Figure~\ref{fig:eos_and_diagnostics}(b) shows that DuplexPO releases suppressed \texttt{<BOS>} impulses more selectively than the SFT model, while Figure~\ref{fig:eos_and_diagnostics}(a) shows consistent \texttt{<EOS>} urge trajectories around user barge-ins. A token-level qualitative comparison with the \emph{SFT Baseline} is provided in Appendix~\ref{app:conv-pattern}. Overall, DuplexPO does not globally suppress speaking intent, but instead shifts behavior near boundary-control decisions, releasing latent \texttt{<BOS>} and \texttt{<EOS>} impulses according to the local conversational state. Consistent with this, Appendix~\ref{app:model-analysis} shows that DuplexPO attends less to the \texttt{<BOS>} sink and more to real-text, recent-context, and PAD-state positions than Moshi 7B.

We further ablate three DuplexPO design choices. First, longer lead times sharply reduce mean RL reward, whereas buffer time has a milder effect, showing that supervision must stay close to boundary events (Appendix~\ref{app:window-ablation}; Figure~\ref{fig:window-ablation}). Second, replacing FCDR with a neural reward model based on a learned temporal state predictor yields partial gains over \emph{SFT Dynamics} but remains less reliable than FCDR, likely because coarse teacher-derived states can disturb fine-grained SFT behaviors. This suggests that explicit FCDR shaping provides a more direct signal for dynamics-critical decisions (Appendix~\ref{app:NRM}). Third, replacing GRPO with a DPO objective based on the highest- and lowest-reward continuations yields weaker performance on most dynamics metrics, particularly yield-oriented metrics, suggesting that group-normalized advantages better exploit the full reward distribution rather than only the extremes (Appendix~\ref{app:opt-ablation}). Additional discussion is provided in Appendix~\ref{app:additional_discussion}.

\paragraph{Limitations}
\label{sec:limitation}
DuplexPO's factorized rewards enable interpretable event-level credit assignment, but may miss subtle pragmatic factors such as user intent, discourse content, speaker style, and culture-specific timing preferences. Human timing annotations are also not unique ground truth, since turn initiation, backchanneling, and yielding can vary across speakers, languages, and conversational settings. As policy updates are restricted to local dynamics-critical windows, DuplexPO may also miss long-range dialogue effects. Future work could combine multilingual and user-adaptive evaluation, and longer-horizon objectives for dialogue coherence.

\section{Conclusion}
We identified the trade-off between conversational dynamics and model intelligence in full-duplex spoken language models. We argue that this trade-off largely stems from coupling what to say with when to speak, rather than from an inherent conflict between the two. We propose DuplexPO, an RL framework that preserves instruction-tuned semantic capability while optimizing real-time speaking decisions over dynamics-critical windows. Experiments show that DuplexPO improves turn-taking, backchanneling, and barge-in handling without degrading instruction following, factual QA, speech understanding, or reasoning.

\newpage
\bibliographystyle{plainnat}
\bibliography{reference}

\newpage
\appendix

\section{Data Details}
\label{app:data-details}
In this chapter, we illustrate all the synthetic and real data we used in this work. In pre-training, we use the speech-continuation data to make the model understand speech input under full-duplex modeling. Then, all other data is used in SFT to improve the reasoning ability and build a helpful speech agent. The broader data-construction recipe follows the released Nemotron-VoiceChat materials~\citep{voicechat}.
\paragraph{1. Speech-continuation data.}
Speech-continuation pre-training uses 530K hours of examples derived from
large text corpora~\citep{su2025nemotron}. Each continuous passage is converted
into a two-party pseudo-dialogue by assigning successive sentences to the user
and agent streams. Turn lengths are randomized: a turn stops after one sentence
with probability $0.8$, while longer turns are formed by appending additional
sentences with a decaying continuation probability; turns longer than 200 words
are handed to the other speaker. The resulting turns are synthesized
with two speaker profiles, aligned in time, and packed into synchronized
two-stream full-duplex examples. We also swap user and agent roles to increase
coverage of both conversational directions.

\paragraph{2. Instruction-following QA data.}
The instruction-following mixture contributes 70K hours of spoken QA data.
Within this set, 10K hours are single-turn examples spanning broad topics, and
the remaining multi-turn examples train the assistant to maintain conversations
lasting up to four minutes. Dialogue generation uses a helpful-assistant
prompt and is conditioned on heterogeneous textual contexts, including
Wikipedia pages.\footnote{\url{https://huggingface.co/datasets/wikimedia/wikipedia}}
The generated turns are then rendered with the multi-speaker TTS pipeline.

\paragraph{3. Synthetic user interruption.}
We augment multi-turn examples with simulated barge-ins. When an agent
utterance lasts more than four seconds, a user interruption is inserted with
probability $0.1$, and its onset is sampled uniformly from the
20\%--80\% span of the agent utterance. After the interruption begins, the
agent stream is given an 8-token reaction delay ($\approx 0.64$\,s) before it
is forced to emit an end-of-sequence (\texttt{<EOS>}) token. At inference time, an \texttt{<EOS>} on
the text channel masks the corresponding audio channel to silence.

\paragraph{4. ASR-QA data.}
ASR-QA examples expose the model to real recording conditions. We use speech
segments from the open-source corpora aggregated in NeMo
ASRSET~\citep{noroozi2024stateful} as acoustic contexts. Following
\citet{noroozi2024instruction}, an LLM writes questions grounded in the ASR
transcripts. These clips are shorter and less knowledge-dense than the
synthetic QA conversations, but they add real speaker variation, channel
effects, and background noise to the full-duplex training mixture.

\paragraph{Model diversity and prompt speaker pool.}
The synthetic portions of our training mixture are generated with several text and acoustic models rather than a single generator. Text responses are sampled from a pool of instruction-tuned LLMs, including GPT-OSS-120B,\footnote{\url{https://huggingface.co/openai/gpt-oss-120b}}
Qwen2.5-72B Instruct,\footnote{\url{https://huggingface.co/Qwen/Qwen2.5-72B-Instruct}} and Llama3.1-70B-Instruct.\footnote{\url{https://huggingface.co/meta-llama/Llama-3.1-70B-Instruct}} Speech rendering is performed with multiple TTS backends, including Chatterbox,\footnote{\url{https://github.com/resemble-ai/chatterbox}} Magpie-TTS,\footnote{\url{https://docs.nvidia.com/nemo-framework/user-guide/latest/speech_ai/magpietts.html}} and Mooncast.\footnote{\url{https://github.com/jzq2000/MoonCast}} For voice conditioning, we build a prompt pool from all 5--10\,s speech segments in LibriTTS~\citep{zen2019libritts}, YODAS~\citep{li2023yodas}, and Hifi-TTS~\citep{bakhturina2021hi}. This pool contains more than 100K prompt segments and covers over 20K speakers.

\paragraph{Dynamics-aware dialogue reconstruction.}
Fisher~\citep{cieri2004fisher} and Seamless-Naturalistic-HQ
(Seamless)~\citep{agrawal2025seamless} provide the natural timing signals used
for rhythm-aware training. Starting from word-level alignments, we recover
utterance segments by merging nearby words while keeping short lexical
acknowledgments as standalone backchannels. For Seamless, we remove
non-dialogue sessions and low-quality utterance fragments before sampling. The
processed training set contains 24.6K Fisher samples and 43.1K Seamless
samples. Splits are fixed at the conversation or session level before
reconstruction, so no evaluation conversation contributes SFT examples, RL
windows, reward targets, or timing annotations. Fisher evaluation uses a
held-out Fisher test SHAR view, while Seamless evaluation uses a held-out
eval-clean view rather than the naturalistic training split. During training, a
randomized round-robin sampler balances the two sources, and each sample is
truncated at 210 seconds.

\paragraph{Acoustic augmentation.}
Acoustic robustness is improved with both feature-level and waveform-level
augmentation. User speech features are augmented with SpecAugment, and the
waveform stream is mixed with a curated set of 10K noise clips from
Freesound~\citep{fonseca2017freesound} and MUSAN~\citep{snyder2015musan}. Noise
is added with probability $0.5$, with the signal-to-noise ratio sampled
uniformly between $0$\,dB and $60$\,dB.

\section{Training Details}
\label{app:training-details}

\paragraph{Implementation.}
The implementation builds on the released Nemotron-VoiceChat
recipe~\citep{voicechat} and is trained with the NeMo Toolkit~\citep{nemo}. We select a smaller language backbone Qwen2.5-7B-Instruct~\citep{qwen2.5} than Nemotron-VoiceChat. Streaming speech is
encoded by a 600M-parameter Parakeet-based encoder with causal convolutional
context,\footnote{\url{https://huggingface.co/nvidia/nemotron-speech-streaming-en-0.6b}}
followed by a 1024-dimensional Transformer modality adapter that maps acoustic
features into the LLM embedding space~\citep{speechencoder1,speechencoder2}.
The speech codec and the streaming flow-matching generator follow
CosyVoice2~\citep{cosyvoice2}. The speech encoder and codec remain frozen
during training. The speech encoder and LLM advance at $12.5$\,Hz, whereas the
audio codec runs at $25$\,Hz; each LLM frame therefore predicts two speech
tokens to keep the text and audio streams temporally aligned. All runs use 64 A800 80\,GB GPUs.

\paragraph{Supervised fine-tuning.}
SFT is applied after speech-continuation pre-training to adapt the model to
assistant-style spoken interaction. The base SFT mixture uses the
instruction-following QA data in Appendix~\ref{app:data-details}, together with ASR-QA and synthetic-interruption examples that expose the model to real
acoustics and basic barge-in handling. \emph{SFT Baseline} uses
this mixture and excludes dynamics-aware dialogue data. \emph{SFT Dynamics}
keeps the same schedule but adds the reconstructed dynamics-aware dialogue data from Section~\ref{sec:dynamic-data}, making it the data-matched SFT
control for DuplexPO. 

\paragraph{Optimization.}
Pre-training, SFT, and RL use AdamW. For pre-training and SFT, we use
$\beta=(0.9,0.98)$, a weight decay of $0$, and an inverse-square-root
learning-rate schedule. Pre-training uses a peak learning rate of
$5\times10^{-4}$ after a 2{,}500-step warm-up, while the SFT stages use a peak
learning rate of $5\times10^{-5}$. These stages are trained with bfloat16
mixed precision, and gradients are clipped to a maximum norm of $1.0$.
Additional RL and reward hyperparameters are summarized in
Table~\ref{tab:rl-reward-hparams}.

\begin{table*}[t]
\centering
\caption{DuplexPO RL and reward hyperparameters used for the reported run. Time values are in seconds unless otherwise noted.}
\label{tab:rl-reward-hparams}
\small
\renewcommand{\arraystretch}{0.92}
\begin{tabularx}{\textwidth}{lllX}
\toprule
Group & Hyperparameter & Value & Role \\
\midrule
Windowing & Frame duration $\Delta$ & 0.08 & LLM decision grid. \\
Windowing & Training lead time $L$ & 1.0 & Context before annotated agent onset. \\
Windowing & Training buffer $B$ & 2.0 & Rollout region after annotated agent offset. \\
Windowing & Validation lead / buffer & 2.0 / 2.0 & Wider evaluation context for testing early starts and yielding behavior. \\
Windowing & Full-turn windows per conversation & 3 & Maximum sampled non-backchannel segments. \\
Windowing & Backchannel windows per conversation & 1 & Maximum sampled backchannel segments. \\
Windowing & Full-turn windows per conversation & 3 & Maximum number of sampled non-backchannel windows. \\
Windowing & Backchannel windows per conversation & 1 & Maximum number of sampled backchannel windows. \\
\midrule
Rollout & Samples per window & 4 & GRPO group size for reward normalization. \\
Rollout & Temperature / top-$p$ & 1.0 / 0.9 & Sampling policy for continuations. \\
Rollout & Maximum rollout steps & 200 & Maximum generated frames inside a window. \\
Rollout & Rollout chunk size & 128 & Chunk size for rollout processing. \\
Optimization & RL learning rate & $1\times10^{-5}$ & AdamW learning rate. \\
Optimization & Warm-up / minimum LR & 100 / $1\times10^{-6}$ & Inverse-square-root schedule. \\
Optimization & KL coefficient $\beta$ & 0.2 & Regularization strength toward the reference policy. \\
Optimization & Advantage clipping & $[-5,5]$ & Clip range for group-normalized advantages. \\
\midrule
Reward & Missed-event penalty & $-0.5$ & Penalty for failing to initiate a target event. \\
Reward & False-alarm penalty & $-0.5$ & Penalty for unwarranted starts. \\
Reward & Backchannel overlong penalty & $-0.5$ & Penalty for floor-grabbing backchannels. \\
Reward & No-\texttt{<EOS>} penalty scale & 0.75 & Scale for failure to stop after user takeover. \\
Reward & Observable margin & 0.16 & Margin for determining observable turn-end evidence. \\
Reward & Interrupt grace & 0.16 & Grace period around user interruption. \\
Reward & Stop tolerance & 0.24 & Allowed delay for yielding after target stop time. \\
Reward & Explicit \texttt{<EOS>} bonus & 0.05 & Small bonus for explicit yielding. \\
Reward & Near-target PAD penalty & $-0.02$ & Penalty for silence near a target start. \\
Reward & Early / late stop penalties & $-0.15$ / $-0.10$ & Penalties for mistimed turn offset. \\
\bottomrule
\end{tabularx}
\end{table*}

\section{Definition of Suppressed Intent Rate (SIR) and Suppression Release Ratio (SRR)}
\label{app:sir-srr}

To quantify a full-duplex agent's ability to \emph{start} speaking when appropriate and to \emph{stop} speaking when the user takes the floor, we introduce two complementary metrics. Both are computed at the frame level on the agent's text head and are defined symmetrically for the two control actions $a^\star \in \{\texttt{<BOS>}, \texttt{<EOS>}\}$.

Let $\pi_{\text{SFT}}$ and $\pi_{\text{RL}}$ denote the SFT and DuplexPO policies. Let $p^{\text{SFT}}_t(a)$ and $p^{\text{RL}}_t(a)$ denote the probability each policy assigns to token $a$ at frame $t$. To isolate differences attributable to the policy head rather than to trajectory divergence, both quantities are evaluated on a single shared context, namely the SFT-generated trajectory. Let $\hat{y}^{\text{SFT}}_t$ and $\hat{y}^{\text{RL}}_t$ denote the corresponding $\arg\max$ tokens. We say that the SFT policy has an \emph{intent} for action $a^\star$ at frame $t$ when $p^{\text{SFT}}_t(a^\star) > \tau$, with $\tau$ a fixed threshold (we use $\tau = 0.1$ throughout).

For each action we restrict attention to a frame set $\mathcal{F}_{a^\star}$ in which the corresponding intent is meaningful. We define $\mathcal{F}_{\texttt{<BOS>}} = \{t \mid \hat{y}^{\text{SFT}}_t = \textsc{pad}\}$ as the set of frames at which the SFT model remains silent, where the intent is the impulse to \emph{start} speaking. We define $\mathcal{F}_{\texttt{<EOS>}} = \{t \mid \hat{y}^{\text{SFT}}_t \neq \texttt{<EOS>} \text{ and the agent is mid-utterance at } t\}$ as the set of frames at which the SFT model continues speaking, where the intent is the impulse to \emph{stop} speaking.

\paragraph{Suppressed Intent Rate (SIR).}
SIR is the fraction of frames in $\mathcal{F}_{a^\star}$ at which the SFT policy holds a latent intent for $a^\star$ but is overridden by a different action,
\begin{equation}
    \mathrm{SIR}_{a^\star}
    \;=\;
    \frac{\bigl|\{ t \in \mathcal{F}_{a^\star}
    \,\mid\, p^{\text{SFT}}_t(a^\star) > \tau \}\bigr|}
    {\bigl|\mathcal{F}_{a^\star}\bigr|}.
\end{equation}
A higher $\mathrm{SIR}_{a^\star}$ indicates that the SFT model has internalised the relevant action intent at the representational level, yet its decision boundary fails to translate that intent into behaviour.

\paragraph{Suppression Release Ratio (SRR).}
SRR is a paired metric defined on exactly the suppressed-intent frames identified by SIR. Let
\begin{equation}
    \mathcal{S}_{a^\star}
    \;=\;
    \{ t \in \mathcal{F}_{a^\star} \,\mid\, p^{\text{SFT}}_t(a^\star) > \tau \}.
\end{equation}
On this same frame set, SRR measures how often the RL policy actually takes action $a^\star$,
\begin{equation}
    \mathrm{SRR}_{a^\star}
    \;=\;
    \frac{\bigl|\{ t \in \mathcal{S}_{a^\star}
    \,\mid\, \hat{y}^{\text{RL}}_t = a^\star \}\bigr|}
    {\bigl|\mathcal{S}_{a^\star}\bigr|}.
\end{equation}
A higher $\mathrm{SRR}_{a^\star}$ indicates that RL's behavioural change is concentrated precisely on frames where the SFT model already exhibited a latent intent, rather than reflecting an indiscriminate shift toward more action-taking.

\section{Attention Analysis}
\label{app:model-analysis}

\begin{figure}[h]
    \centering
    \includegraphics[width=\linewidth]{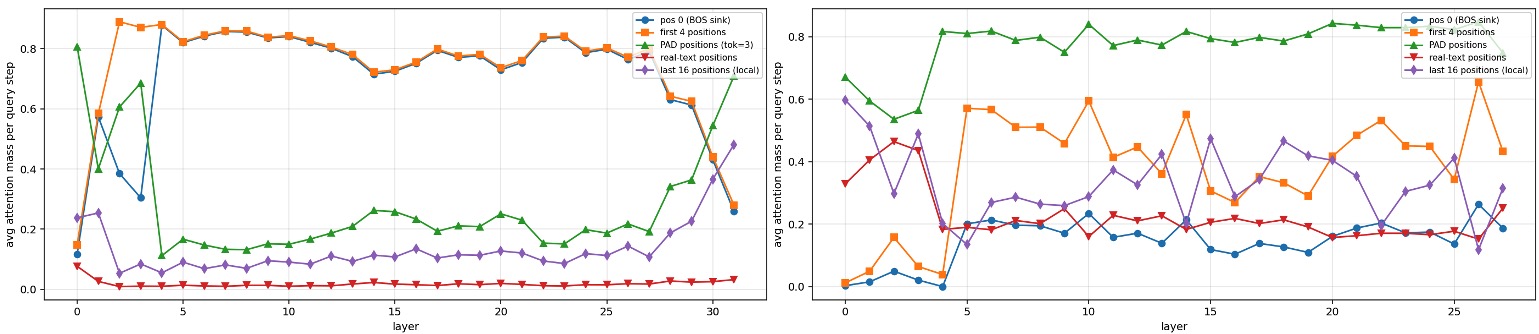}
    \caption{Attention mass distribution across token categories. Attention mass is averaged across layers and grouped by token category, including the \texttt{<BOS>} token, early positions, PAD positions, real-text tokens, and recent-context tokens. In our streaming representation, PAD positions encode non-speech or waiting states that can inform speaking and yielding decisions. Compared with Moshi 7B, DuplexPO assigns less mass to the initial \texttt{<BOS>} sink and more mass to real-text, recent-context, and PAD-state positions.}
    \label{fig:attention_mass}
\end{figure}

This appendix provides additional analysis for the model behaviour
analysed in Section~\ref{sec:evaluation}. The attention statistics in Figure~\ref{fig:attention_mass} and Table~\ref{tab:attention} should be read as supporting evidence about context usage, not as a mechanistic proof of the learned timing policy.

\begin{table}[h]
\centering
\caption{Attention mass distribution (\%) across token categories.}
\label{tab:attention}
\small
\begin{tabular}{lcc}
\toprule
Metric & Moshi 7B~\citep{moshi} & DuplexPO \\
\midrule
\texttt{<BOS>} sink (pos.\ 0)      & 70.2 & 14.5 \\
First 4 positions       & 74.2 & 37.9 \\
PAD positions           & 28.0 & 77.5 \\
Real-text positions     &  1.8 & 22.5 \\
Last 16 tokens          & 13.5 & 33.4 \\
Attention entropy       & 1.63 &  3.19 \\
\bottomrule
\end{tabular}
\end{table}

\section{Judge Calibration with Synthetic Timing Perturbations}
\label{app:judge_calibration}

To assess whether the pairwise LLM judge is sensitive to conversational timing rather than transcript content alone, we conduct a controlled timing-perturbation sanity check on Fisher. For each conversation, we keep the user timeline and the agent transcript unchanged, but shift the agent timestamps by $\pm 1.0$ seconds. A $+1.0$s shift makes agent responses and backchannels systematically late, while a $-1.0$s shift makes agent speech more likely to occur prematurely during ongoing user speech. The judge is then asked to compare the original timeline with the perturbed timeline using the same blinded pairwise protocol as in Appendix~\ref{app:llm-pairwise-judge}. Candidate order is randomized across examples.

Importantly, in this calibration experiment, we remove aggregate dynamics statistics from the judge input. Thus, the judge must rely on the aligned timestamps and transcripts rather than precomputed automatic metrics. Win rates are computed over non-tie judgments, and statistical significance is evaluated with a two-sided binomial sign test.

As shown in Table~\ref{tab:judge_calibration_overall}, the judge prefers the original timing over both delayed and advanced perturbations. The effect is significant for advanced perturbations and marginal for delayed ones. Because the transcript content is identical across candidates, the results indicate a timing preference aligned with human conversational perception, with stronger sensitivity to premature turn entry than to comparable response delays \citep{sacks1974simplest, stivers2009universals, levinson2015timing}. This asymmetry also supports the design of FCDR’s onset reward, which applies a narrower tolerance to early initiation than to delayed responses.

\begin{table}[t]
\centering
\small
\caption{
Judge calibration under synthetic timing perturbations on Fisher. The transcript content is identical across candidates; only the agent timestamps are shifted. Candidate order is randomized. Win rates are computed over non-tie judgments. $p$-values are computed using a two-sided binomial sign test. 
}
\label{tab:judge_calibration_overall}
\begin{tabular}{lcccccc}
\toprule
Perturbation & \# Conv. & Original Wins & Perturbed Wins & Ties & Original Win Rate & $p$-value \\
\midrule
Delayed by $+1.0$s  & 26 & 18 & 8 & 0 & 69.2\% & 0.076 \\
Advanced by $-1.0$s & 26 & 21 & 5 & 0 & 80.8\% & 0.002\\
\bottomrule
\end{tabular}
\end{table}
\section{LLM Pairwise Judge for Full-Duplex Dialogue Naturalness}
\label{app:llm-pairwise-judge}

The pairwise judge is designed to evaluate conversational dynamics rather
than semantic answer quality. Table~\ref{tab:judge-system-prompt} gives
the system instruction, and Table~\ref{tab:judge-user-prompt} gives the
user-prompt template used to present aligned candidate timelines.
We used Gemini 3.0 Pro and counted a result only when the returned JSON
matched the requested schema and contained no error. Candidate labels were
anonymized in the prompt, and in these runs candidate A is the \emph{SFT Baseline} and candidate B is DuplexPO. Table~\ref{tab:judge-summary} reports both overall and dimension-level preferences. Win rates are computed over non-tie judgments, and the $p$-values use a two-sided binomial sign test over the non-tie preferences.
\begin{table}[h]
\centering
\caption{Conversation-level LLM pairwise judge results. Candidate B is
DuplexPO; candidate A is the \emph{SFT Baseline}. Win rates exclude ties.}
\label{tab:judge-summary}
\small
\renewcommand{\arraystretch}{0.9}
\resizebox{\linewidth}{!}{
\begin{tabular}{llrrrrrr}
\toprule
Dataset & Criterion & Total & B wins & A wins & Ties & B win rate & $p$-value \\
\midrule
Fisher & Overall & 26 & 20 & 6 & 0 & 76.9 & 0.009 \\
Fisher & Turn-taking & 26 & 20 & 6 & 0 & 76.9 & 0.009 \\
Fisher & Backchannel & 26 & 22 & 2 & 2 & 91.7 & $<10^{-4}$ \\
Fisher & Barge-in handling & 26 & 21 & 3 & 2 & 87.5 & $<10^{-3}$ \\
\midrule
Seamless & Overall & 383 & 264 & 117 & 2 & 69.3 & $<10^{-13}$ \\
Seamless & Turn-taking & 383 & 250 & 123 & 10 & 67.0 & $<10^{-10}$ \\
Seamless & Backchannel & 383 & 261 & 54 & 68 & 82.9 & $<10^{-32}$ \\
Seamless & Barge-in handling & 383 & 237 & 70 & 76 & 77.2 & $<10^{-21}$ \\
\bottomrule
\end{tabular}
}
\end{table}

\begin{table}[p]
\centering
\caption{System prompt for LLM pairwise judging of full-duplex dialogue
dynamics.}
\label{tab:judge-system-prompt}
\begin{tabular}{|p{0.96\linewidth}|}
\hline
\textbf{System prompt} \\
\hline
\begin{Verbatim}[breaklines=true,breakanywhere=true,fontsize=\footnotesize]
You are an expert evaluator for full-duplex spoken dialogue rhythm.

Your role:
- Compare two candidate system turns from the same local dialogue context.
- Evaluate ONLY three interaction abilities: turn-taking, backchanneling,
  and user barge-in handling.
- Use only the observable local dialogue context and candidate turns
  provided in the prompt. Do NOT assume access to any hidden annotations,
  target response, target timing interval, or dataset label.
- Do NOT evaluate factual correctness, knowledge, helpfulness,
  informativeness, or whether the answer fully solves the user's question.
- Do NOT prefer a candidate merely because it is more detailed, more
  informative, or more semantically complete.
- If both candidates are similar on these three interaction abilities,
  output "tie".

Evaluation dimensions:
1. turn_taking: Which candidate starts at a more natural time and yields
   smoothly for the local conversation state. Do not reward a candidate
   for being longer or more substantive.
2. backchannel: Which candidate better uses short listener feedback when
   appropriate, without being stale, too long, or floor-grabbing.
3. user_barge_in_handling: Which candidate better avoids talking over
   the user or better yields/gets out of the way when the user starts or
   continues speaking.

Important judgment rules:
- Judge from timing and local dialogue behaviour, not answer quality.
- A short response can be better if it functions as a natural backchannel.
- A longer response should not be preferred just because it holds the
  floor or gives more content.
- Penalise competitive interruption, awkward delayed feedback, excessive
  overlap with the user, and failure to yield when the user appears to
  take the floor.
- If one dimension is hard to determine from the visible context and
  candidate timestamps, mark that dimension as "tie".

You MUST:
1. Read the shared local context first.
2. Infer the local turn-taking state: user_holding_floor,
   user_yielding_floor, backchannel_opportunity, full_turn_opportunity,
   or uncertain.
3. Compare candidate A and candidate B on the three dimensions.
4. State preferred_candidate as "A", "B", or "tie".
5. Give concise reasons focused only on turn-taking, backchanneling,
   and user barge-in handling.
6. Give confidence from 0 to 1.

Output MUST be valid JSON matching this schema, with no markdown and no
extra text:
{
  "interaction_id": "string",
  "pair_id": "string",
  "candidate_a_id": "string",
  "candidate_b_id": "string",
  "inferred_turn_state": "user_holding_floor|user_yielding_floor|
                          backchannel_opportunity|full_turn_opportunity|
                          uncertain",
  "preferred_candidate": "A|B|tie",
  "preference_reason": "string",
  "dimension_preferences": {
    "turn_taking":            {"preferred_candidate": "A|B|tie",
                               "reason": "string"},
    "backchannel":            {"preferred_candidate": "A|B|tie",
                               "reason": "string"},
    "user_barge_in_handling": {"preferred_candidate": "A|B|tie",
                               "reason": "string"}
  },
  "major_rhythm_issues": ["string"],
  "content_quality_ignored_note": "string",
  "confidence": 0.0-1.0
}
\end{Verbatim}
\\
\hline
\end{tabular}
\end{table}

\begin{table}[p]
\centering
\caption{User-prompt template for conversation-level pairwise dynamics
judging.}
\label{tab:judge-user-prompt}
\begin{tabular}{|p{0.96\linewidth}|}
\hline
\textbf{User prompt template} \\
\hline
\begin{Verbatim}[breaklines=true,breakanywhere=true,fontsize=\footnotesize]
Compare candidate A and candidate B for overall full-duplex
spoken-dialogue rhythm across this conversation sample. Ignore factual
correctness, answer completeness, and informativeness. Output only valid
JSON matching the schema.

## Interaction ID
{interaction_id}

## Shared user timeline
- [user] ({user_turn_1_start_s:.2f}s--{user_turn_1_end_s:.2f}s)
  {user_turn_1_text}
- [user] ({user_turn_2_start_s:.2f}s--{user_turn_2_end_s:.2f}s)
  {user_turn_2_text}
...

## Candidate A assistant timeline
- [candidate] ({candidate_a_turn_1_start_s:.2f}s--
                {candidate_a_turn_1_end_s:.2f}s)
  {candidate_a_turn_1_text}
- [candidate] ({candidate_a_turn_2_start_s:.2f}s--
                {candidate_a_turn_2_end_s:.2f}s)
  {candidate_a_turn_2_text}
...

## Candidate A observable rhythm summary
- n_turns:                      {candidate_a_n_turns}
- total_speech_ms:              {candidate_a_total_speech_ms}
- total_user_overlap_ms:        {candidate_a_total_user_overlap_ms}
- n_user_overlaps:              {candidate_a_n_user_overlaps}
- n_interrupted_by_user:        {candidate_a_n_interrupted_by_user}
- median_pause_after_user_ms:   {candidate_a_median_pause_after_user_ms}

## Candidate B assistant timeline
- [candidate] ({candidate_b_turn_1_start_s:.2f}s--
                {candidate_b_turn_1_end_s:.2f}s)
  {candidate_b_turn_1_text}
- [candidate] ({candidate_b_turn_2_start_s:.2f}s--
                {candidate_b_turn_2_end_s:.2f}s)
  {candidate_b_turn_2_text}
...

## Candidate B observable rhythm summary
- n_turns:                      {candidate_b_n_turns}
- total_speech_ms:              {candidate_b_total_speech_ms}
- total_user_overlap_ms:        {candidate_b_total_user_overlap_ms}
- n_user_overlaps:              {candidate_b_n_user_overlaps}
- n_interrupted_by_user:        {candidate_b_n_interrupted_by_user}
- median_pause_after_user_ms:   {candidate_b_median_pause_after_user_ms}
\end{Verbatim}
\\
\hline
\end{tabular}
\end{table}

\section{Reward Visualisation}
\label{app:reward_vis}

\begin{figure}[t]
  \centering
  \includegraphics[width=\linewidth]{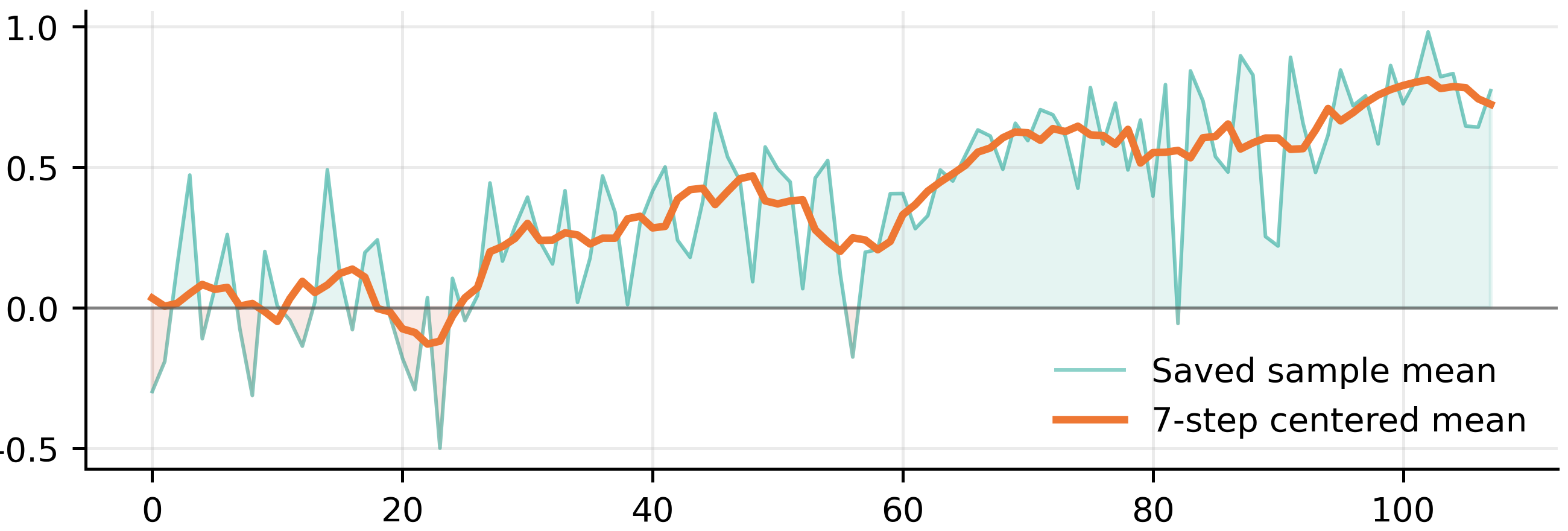}
  \caption{Training dynamics of the total reward and its individual
  components across optimisation steps. The total reward rises
  monotonically and then plateaus, indicating stable optimisation.}
  \label{fig:reward_curve_dynamics}
\end{figure}

\begin{figure}[t]
  \centering
  \includegraphics[width=\linewidth]{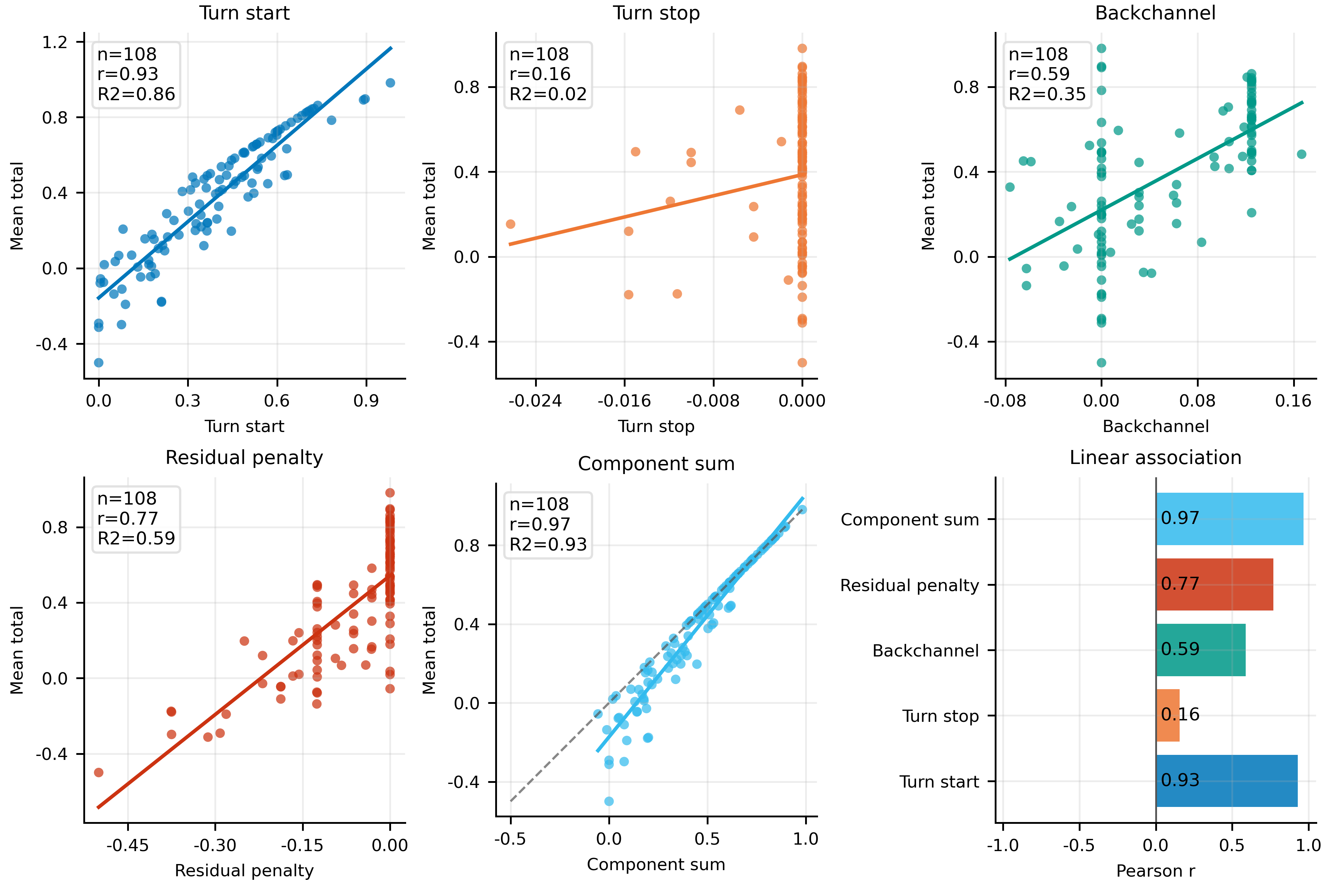}
  \caption{Pairwise correlations between reward components and the
  behavioural events they shape. Stop and start rewards specialise on
  yielding and turn initiation respectively, with low cross-talk.}
  \label{fig:reward_component_relationships}
\end{figure}

As shown in Figure~\ref{fig:reward_curve_dynamics}, the reward curve
increases monotonically and then stabilises, suggesting that training is
robust and generalises across dialogue contexts. The pairwise
correlation analysis in Figure~\ref{fig:reward_component_relationships} highlights
the specialisation of reward signals: the stop reward provides the
strongest signal specifically for backchannel yielding, while start
rewards drive the timing of turn initiation. These correlations
indicate that the model decomposes conversational dynamics into distinct,
actionable components, effectively addressing both \emph{when to yield}
and \emph{when to participate}.

\section{Ablation Study: Effect of Window Boundary Size}
\label{app:window-ablation}
\begin{figure}
    \centering
    \small
    \includegraphics[width=0.8\linewidth]{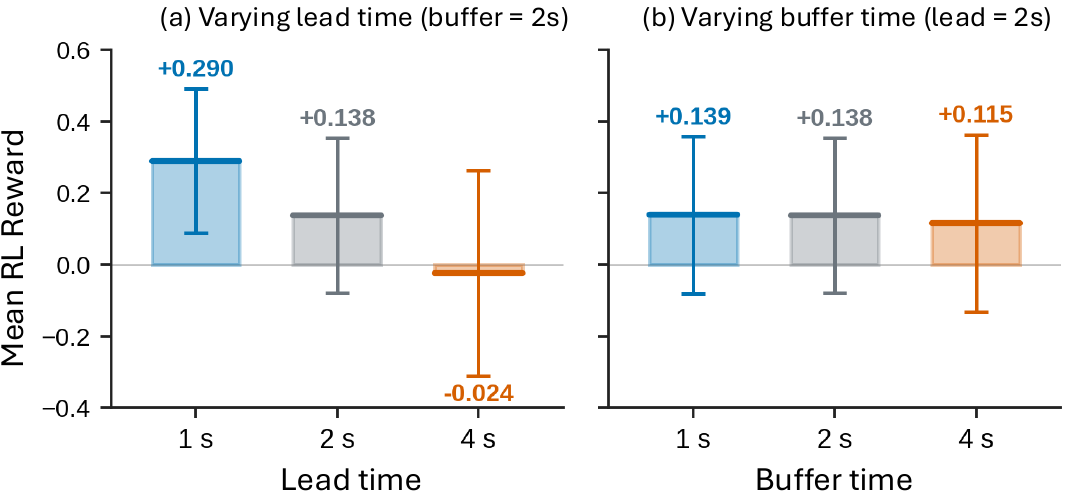}
    \caption{Ablation of dynamics-critical window lead and buffer times. Bars show the mean RL reward across checkpoints every 10 training steps. Error bars show $\pm 1$ standard deviation across checkpoints.}
    \label{fig:window-ablation}
\end{figure}

We separately ablate window lead time $L$ and window buffer time $B$, as shown in Figure~\ref{fig:window-ablation}. The two parameters affect the RL reward differently. Varying Window Lead Time produces a clear ordering in mean reward, with overly long lead times causing a substantial performance drop. In contrast, varying Window Buffer Time yields no significant separation, with the mean rewards differing by at most $0.02$. These results suggest that dynamics-critical window sampling is primarily sensitive to the amount of anticipatory context required for committing to a speaking decision, rather than to the amount of post-event context retained in the window.

\section{Ablation Study: Neural Reward Model}
\label{app:NRM}

The neural reward model teacher is SoulX-Duplug \citep{yan2026soulx}, denoted \(f_\phi\). This method does not use assistant ground-truth. For each training sample, the user waveform \(u\) and its aligned word sequence \(\{(w_j,t_j^{s},t_j^{e},t_j^{emit})\}_j\) are fed to \(f_\phi\), producing a chunk-level teacher timeline
\begin{equation}
\mathcal{T}=\{(a_m,b_m,s_m)\}_{m=1}^{M},\qquad
s_m\in\{\texttt{idle},\texttt{nonidle},\texttt{speak},\texttt{blank}\}.
\end{equation}
The neural reward model first converts the SoulX-Duplug timeline into a state \(s_t\) for each frame. A prescription prior \(p_T(d\mid s,c)\) scores whether the agent decision \(d_t\) is appropriate under teacher state \(s_t\) and transition type \(c_i\in\{\text{yield},\text{interrupt}\}\). For a predicted sequence \(d\) in window \(W_i\), we compute the salience-weighted teacher score
\begin{equation}
A_i(d)=
\frac{\sum_{t=1}^{n_i} w_{i,t}\log p_T(d_t\mid s_{i,t},c_i)}
{\sum_{t=1}^{n_i} w_{i,t}},
\end{equation}
where blank frames have $w_{i,t}=0$, frames within $\sim$160\,ms of a SoulX-Duplug transition are weighted $1.0$ and other non-blank frames $0.5$. The core reward subtracts a per-window frozen-reference baseline:
\begin{equation}
r^{\mathrm{core}}_{i,k}
=
A_i(g(z_{i,k}))-
\frac{1}{K_0}\sum_{l=1}^{K_0} A_i(g(z^{\mathrm{ref}}_{i,l})).
\end{equation}

\begin{table}[h]
\centering
\caption{Comparison with SFT controls and reward calculation ablation. 
SFT Baseline and SFT Dynamics are non-RL controls without reward optimization. 
NRM and FCDR use the same dynamics-critical window sampler and GRPO-style optimization, 
and differ only in the reward calculation method. Rates are in \%; Onset MAE is in seconds. 
Best values in each column are bolded.}
\label{tab:reward-ablation}
\small
\renewcommand{\arraystretch}{0.8}
\setlength{\tabcolsep}{4pt}
\begin{tabular}{llccccccc}
\toprule
\multirow{2}{*}{Dataset} & \multirow{2}{*}{Training setting} & \multirow{2}{*}{Onset MAE ($\downarrow$)}
& \multicolumn{2}{c}{Turn-taking}
& \multicolumn{2}{c}{Backchannel}
& \multicolumn{2}{c}{Barge-in}\\
\cmidrule(lr){4-5}\cmidrule(lr){6-7}\cmidrule(lr){8-9}
 & & & Init ($\uparrow$) & Yield ($\uparrow$)
 & Init ($\uparrow$) & Yield ($\uparrow$)
 & VIR ($\downarrow$) & Yield ($\uparrow$)\\
\midrule

\multirow{4}{*}{Fisher}
& \emph{SFT Baseline} & 1.22 & 91.8 & 78.5 & 92.2 & 79.4 & -- & --\\
& \emph{SFT Dynamics} & 1.34 & 79.7 & 63.0 & 83.2 & 63.4 & -- & --\\
& NRM reward  & 1.22 & 96.4 & 81.6 & 89.2 & 50.0 & -- & -- \\
& FCDR reward & \textbf{0.69} & \textbf{100.0} & \textbf{98.7} & \textbf{97.8} & \textbf{100.0} & -- & -- \\

\midrule
\multirow{4}{*}{Seamless}
& \emph{SFT Baseline} & 1.22 & 91.8 & 78.5 & 92.2 & 79.4 & -- & --\\
& \emph{SFT Dynamics} & 1.34 & 79.7 & 63.0 & 83.2 & 63.4 & -- & --\\
& NRM reward  & 1.32 & 84.5 & 67.6 & 93.5 & 74.0 & -- & -- \\
& FCDR reward & \textbf{1.03} & \textbf{98.0} & \textbf{93.6} & \textbf{99.5} & \textbf{93.3} & -- & -- \\

\midrule
\multirow{4}{*}{FDB-v3}
& \emph{SFT Baseline} & -- & -- & -- & -- & -- & 8.0 & 64.8 \\
& \emph{SFT Dynamics} & -- & -- & -- & -- & -- & \textbf{4.0} & 93.3\\
& NRM reward  & -- & -- & -- & -- & -- & \textbf{4.0} & 93.3 \\
& FCDR reward & -- & -- & -- & -- & -- & 5.0 & \textbf{100.0} \\

\bottomrule
\end{tabular}
\end{table}
Table~\ref{tab:reward-ablation} shows that NRM provides a useful but less reliable dynamics signal than FCDR. Although NRM improves over \emph{SFT Dynamics} on several metrics, it does not consistently surpass the original \emph{SFT Baseline}. This suggests that teacher-derived state supervision can recover some temporal coordination cues, but may also disturb behaviors already learned during SFT, especially for fine-grained decisions such as backchannel yielding and ambiguous turn transitions. One likely reason is that the neural teacher provides coarse temporal states rather than direct feedback on whether a specific \texttt{<BOS>} or \texttt{<EOS>} decision is appropriate within the current dynamics-critical window. In contrast, FCDR directly rewards target floor-control decisions, leading to more stable improvements across turn-taking, backchanneling, and barge-in handling.

\section{Ablation Study: Optimization Methods}
\label{app:opt-ablation}

Holding the FCDR-based reward calculation and the dynamics-critical window sampler fixed, we ablate the preference-optimization objective by comparing GRPO with a DPO-style variant. In the DPO-style variant, the highest- and lowest-reward continuations within each sampled window are used as the preferred and dispreferred pair. Table~\ref{tab:opt-ablation} reports the resulting dynamics metrics on Fisher, Seamless, and FDB-v3.

\begin{table}[h]
\centering
\caption{Optimization method ablation. Both variants use the same FCDR and dynamics-critical window sampler, and differ only in the preference-optimization objective. Rates are reported in \%; Onset MAE is reported in seconds. The best result in each column is bolded.}
\label{tab:opt-ablation}
\small
\renewcommand{\arraystretch}{0.8}
\setlength{\tabcolsep}{4pt}
\begin{tabular}{llccccccc}
\toprule
\multirow{2}{*}{Dataset} & \multirow{2}{*}{Optimization} & \multirow{2}{*}{Onset MAE ($\downarrow$)}
& \multicolumn{2}{c}{Turn-taking}
& \multicolumn{2}{c}{Backchannel}
& \multicolumn{2}{c}{Barge-in}\\
\cmidrule(lr){4-5}\cmidrule(lr){6-7}\cmidrule(lr){8-9}
 & & & Init ($\uparrow$) & Yield ($\uparrow$)
 & Init ($\uparrow$) & Yield ($\uparrow$)
 & VIR ($\downarrow$) & Yield ($\uparrow$)\\
\midrule
\multirow{2}{*}{Fisher}
& DPO  & 0.81 & \textbf{100.0} &  93.6      & 91.4 &        71.4    & -- & -- \\
& GRPO &\textbf{0.69} &\textbf{100.0}&\textbf{98.7}&\textbf{97.8} & \textbf{100.0}   & -- & -- \\
\midrule
\multirow{2}{*}{Seamless}
& DPO  & \textbf{1.00}& 95.1 &    77.8    & 95.8 &84.7  & -- & -- \\
& GRPO & 1.03& \textbf{98.0} & \textbf{93.6} & \textbf{99.5} & \textbf{93.3} & -- & -- \\
\midrule
\multirow{2}{*}{FDB-v3}
& DPO  & -- & -- & -- & -- & --&\textbf{4.0}  & 93.3  \\
& GRPO & -- & -- & -- & -- & -- & 5.0 & \textbf{100.0} \\
\bottomrule
\end{tabular}
\end{table}

The results show that GRPO outperforms the DPO-style objective on most window-level dynamics metrics, especially yield rate and backchannel behavior. This suggests that reducing each window to only the highest- and lowest-reward continuations discards useful ranking information. In contrast, GRPO uses group-normalized advantages over the full set of sampled continuations, providing a denser and more stable optimization signal for dynamics-critical decisions. Although DPO achieves slightly better Onset MAE on Seamless and a lower VIR on FDB-v3, GRPO achieves higher yield rates across all datasets, indicating better recovery and response behavior in dynamics-critical scenarios.
\begin{figure}[t]
  \centering
  \includegraphics[width=0.9\linewidth]{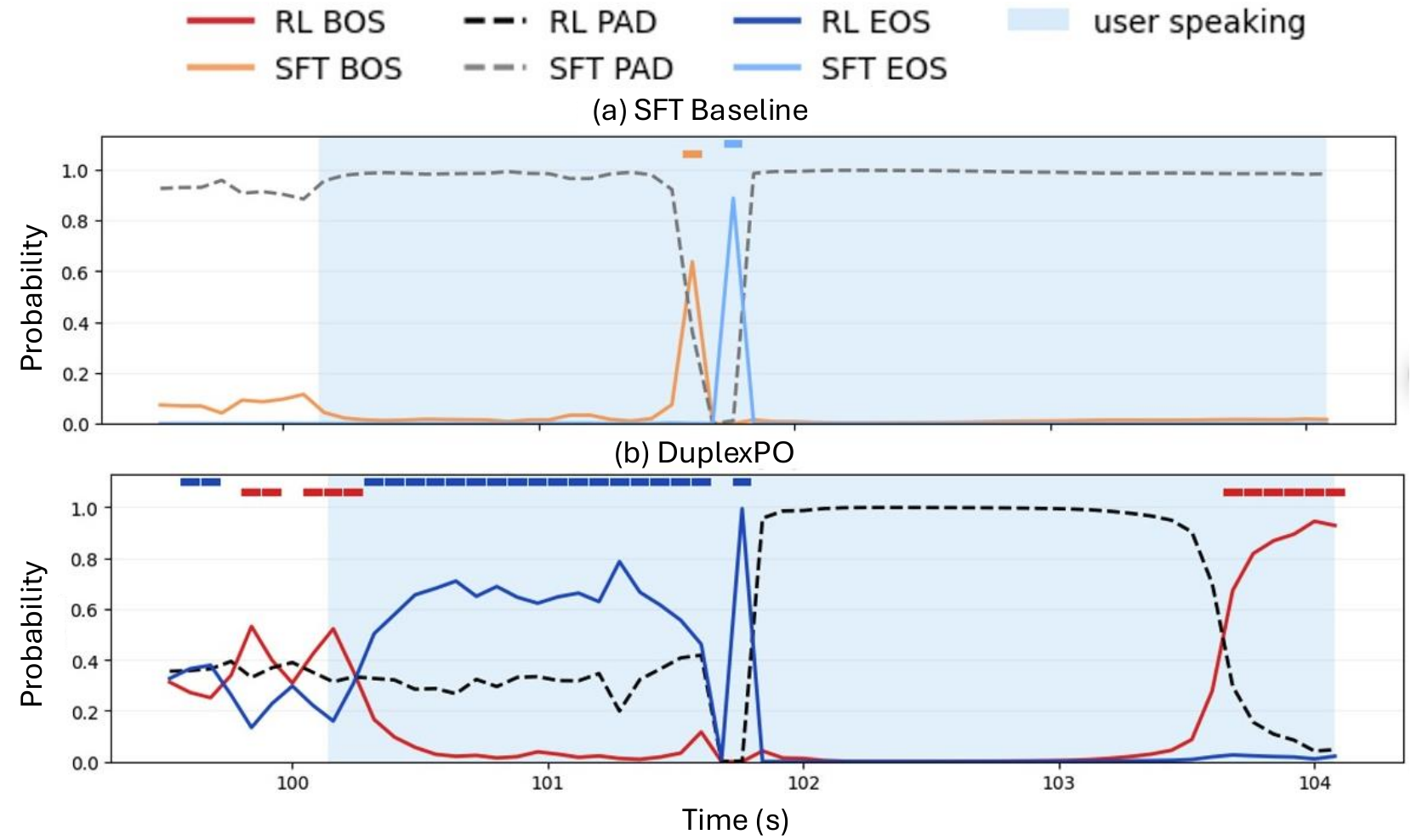}
  \caption{Example token-level conversational pattern. Compared with
  the \emph{SFT baseline}, DuplexPO maintains and releases \texttt{<BOS>}/\texttt{<EOS>}
  impulses around user speech instead of staying near PAD.}
  \label{fig:conv_pattern}
\end{figure}
\section{Example of a Conversational Pattern}
\label{app:conv-pattern}

Figure~\ref{fig:conv_pattern} contrasts the token-level conversational
pattern of DuplexPO with that of the \emph{SFT baseline}.
DuplexPO maintains and releases \texttt{<BOS>}/\texttt{<EOS>} impulses around user
speech instead of remaining near PAD, which is consistent with the
attention statistics in Appendix~\ref{app:model-analysis} and the
suppression metrics defined in Appendix~\ref{app:sir-srr}.

\section{Broader Impacts}
\label{app:broader_impacts}
This work improves the naturalness of full-duplex spoken dialogue models. Positively, this can significantly enhance accessible interfaces and conversational assistants. Negatively, highly human-like voice dynamics (e.g., natural backchanneling and interruption) could be misused for deceptive practices like realistic vishing (voice phishing). Future deployments should consider appropriate safeguards, such as watermarking or identity disclosure, to mitigate these risks.

\clearpage

\end{document}